# SAPNet: a deep learning model for identification of single-molecule peptide post-translational modifications with surface enhanced Raman spectroscopy

Mulusew W.Yaltaye[1,2], Yingqi Zhao[1]*, Eva Bozo [1], Pei-Lin Xin [1], Vahid Farrah [1,3], Francesco De Angelis[4] and Jian-An Huang [1,2,]*

[1] Research Unit of Health Sciences and Technology, Faculty of Medicine, University of Oulu, Aapistie 5 A, 90220 Oulu, Finland.

[2] Research Unit of Disease Networks, Faculty of Biochemistry and Molecular Medicine, University of Oulu, Aapistie 5 A, 90220 Oulu, Finland.

[3] Institute for Sport and Sport Science, TU Dortmund University, Dortmund, Germany.

[4] Istituto Italiano di Tecnologia, Via Morego 30, 16163, Genoa, Italy.

**Abstract**:

Nanopore resistive pulse sensors are emerging technologies for single-molecule protein sequencing. But they can hardly detect small post-translational modifications (PTMs) such as hydroxylation in single-molecule level. While a combination of surface enhanced Raman spectroscopy (SERS) with plasmonic nanopores can detect the small PTMs, the blinking Raman peaks in the single-molecule SERS spectra leads to a big challenge in data analysis and PTM identification. Herein, we developed and validated a one-dimensional convolutional neural network (1D-CNN) for amino acids and peptides identification from their PTMs including hydroxylation and phosphorylation by their single-molecule SERS spectra, named Single Amino acid and Peptide Network (SAPNet). Our work combines cutting-edge plasmonic nanopore technology for SERS signal acquisition and deep learning for fully automated extraction of information from the SERS signals. The SAPNet model achieved an overall accuracy of 99.66% for the identification of amino acids from their modification, and 98.38% for the identification of peptides from their PTM translation. We also evaluated the model with out-of-sample examples with good performance. Our work can be beneficial for early detection of diseases such as cancers and Alzheimer's disease.

**Introduction**

Post-translational modifications (PTMs) play essential roles in protein signaling, function, localization, and other important biological processes. [1] The existence of PTM in the human body may finally cause serious health consequences. [2] Protein PTM analysis of trace amounts of biomarkers in human biofluids is of great significance for early-stage disease diagnosis and low-abundance protein studies. [3] Mass spectrometry is currently the mainm technology for label-free protein analysis, but its sensitivity is limited by the requirement of $10^6$-$10^8$ copies of molecules. [4] PTMs identification using mass spectroscopy also suffers false identification due to the change of the combinatorial rules for spectra explanation, molecules of the same mass-to-charge ratio, and insufficient sensitivity. [5,6]

Nanopore resistive pulse sensing [7,8] is an emerging technology that allows the monitoring of single molecules translocating through the nanopore. By monitoring the current change induced by modification of amino acid residues, PTMs had been successfully detected in nanopores including phosphorylation, [9–11] acetylation, [12] propionylation, [13] glycosylation, [14,15] nitration and oxidation. [16] But it can not detect small PTMs such as hydroxylation, because the hydroxylation can induce so little current drop that the resistive pulses of hydroxylation are hard to be well defined.

Surface-enhanced Raman spectroscopy (SERS) is a label-free highly sensitive molecule detection technology that has been widely used in protein analysis and PTM detection. [17] Intense investigations have been performed on identifying various kinds of PTMs, including acetylation, alkylation, glycation, nitration, and phosphorylation. [18] However, with conventional SERS enhancing structure, for example, gold or silver nanoparticle arrays, the modifications usually cannot induce spectral change which is strong enough for single molecule analysis. Even at the multi-molecule level, the PTM identifications are challenging due to the high noise-to-signal ratio, lack of characteristic peaks, or limited characteristic peak shifts. [19]

Recently, we have developed a particle-in-pore SERS platform (Figure 1a) that demonstrates the capability of single molecule detection of amino acids [20] and DNA bases. [19] Gold particles adsorbed with analytes were flowed through and trapped to the sidewall of the nanopore under optical and electric forces, therefore generating a single plasmonic hot spots with extremely strong enhancement and spatial resolution for single molecule detection. [21] With the capability of generating high-quality single-molecule SERS spectra, the particle-in-pore technique is ideal for detecting modified amino acid residue. In this paper, for the first time, we obtain SERS spectra of single amino acids and peptides before and after phosphorylation and hydroxylation.

Another challenge to achieving PTM identification at a single molecule level using the particle-in-pore system is the spectra information extraction. The single molecule spectra generated in narrow hot spots are inherently vibrating due to the molecule Brownian motion with variating molecule position and orientation in the hot spot. Manual peak assignment in a traditional way was very difficult and time-consuming due to lack of previous literature of single molecule PTM, the complexity of vibrating spectra, as well as the large amount of data generated.

Taking the advantage of large data amount, the synergy of SERS signals with convolutional neural network deep learning analysis gives unprecedented opportunities to extract information from complex spectra. [22–24] The deep convolutional neural network is excellent at feature extraction and pattern recognition tasks, [25] contributing to distinguishing small differences in biomolecular patterns. The hierarchical architecture, composed of repeated convolution and pooling operations, enables the detection of subtle signals across multiple scales. Using such deep learning techniques with automated feature extraction, we can bypass manual peak-by-peak assignment and analysis and extract the data features that indicate the differences between molecules in an automated manner. With further optimization and a huge training dataset, deep learning models offer a promising path toward robust, label-free analysis of individual biomolecules.

In this paper, by combining the particle-in-pore SERS enhancement technique and deep convolutional neural network, we successfully discriminated two typical types of PTM: phosphorylation and hydroxylation in both amino acids and peptides based on their single molecule SERS spectra. Phosphorylation on amino aicds, peptides and protein had been widely studied by various detection technique, such as mass spectrometry [26], nanopore [27–29], and SERS [30,31], due to its importance in biological porcess [32] as well as relatively strong spectrum change incuded by phosphate group. In contrast, the detection of hydroxylation is even more chanllenging because of small molecule weight of hydroxy group which usually generate little signal changes. In this work, taking the advantage of ultra sensitivity of our SERS enhancement technique, we succesfully demonstrated the discrimination of Proline and Hydroxyproline at single-molecule level.

The molecules detected in this work include Proline, Hydroxyproline, Tyrosine, Phosphorylated Tyrosine, Serine, Phosphorylated Serine as well as two peptides derived from Hypoxia-Inducible Factor-1α (HIF-1α) and fetuin-A (FETUA), as shown in Table 1. Both HIF-1α and FETUA were selected as model compounds for hydroxylation and phosphorylation respectively because they are important biomarkers and their PTM has been well studied by mass spectometery. [33–35] The hydroxylation of conserved prolyl residues in the central region of HIF-1α promotes molecule interactions, which targets HIF-1α for degradation by the ubiquitin–proteasome pathway.[36,37] peptided with same sequece of proline residue containing site of HIF-1α was used for discrimination of hydroxylation (See Table 1 for the details). As a multifunctional peptide which plays an important role in diabetes, kidney disease, and cancer, [38] monitoring FETUA's phoysphoration can be beneficial to dignosis and treatment.[39] A peptide contain the same sequence with the phosphoralation cite was used for PTM discrimination (See Table 1 for the details). We achieved an overall accuracy of 99.66% for the identification of amino acids from their modification, and 98.38% for the identification of peptides from their PTM mixtures. [40,41] This is asignificant step towards single-molecule protein analysis and sequencing, which has a vast application in the almost all biomedical fields.

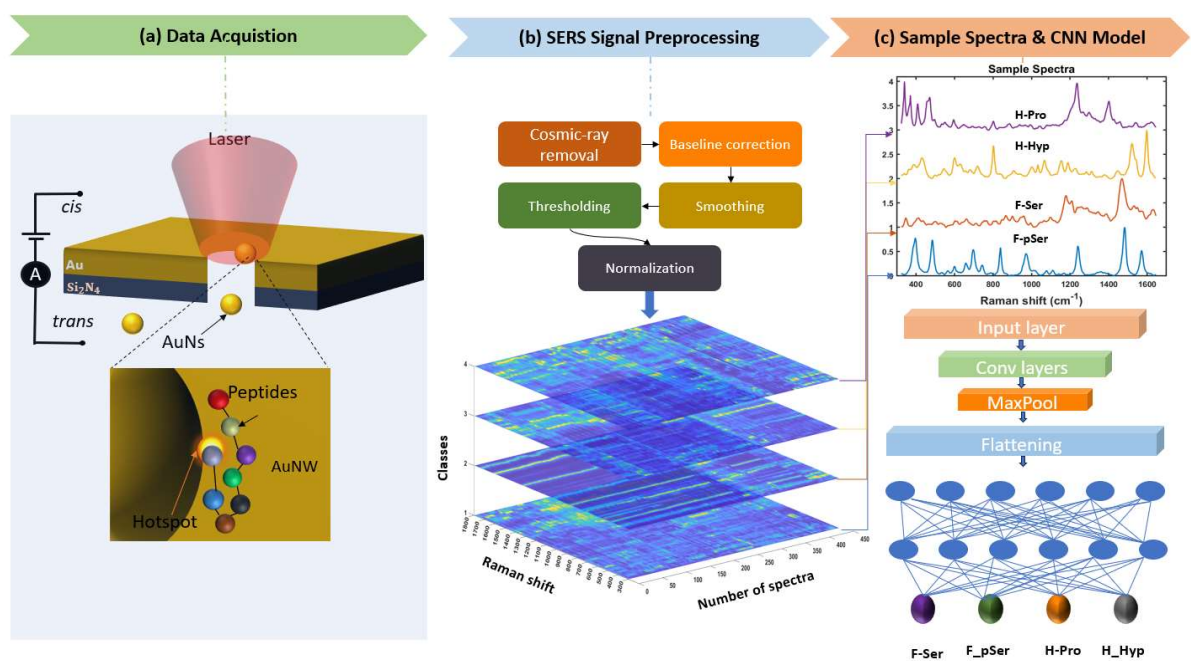

*Figure 1.* Workflow of amino acid and peptide identification from the SERS signal obtained from a plasmonic nanopore using a deep convolutional neural network. (a) in the first half, the schematic of electro-plasmonic trapping mechanism in which a single gold nanosphere (AuNs) is trapped in the gold nanohole and excited with CW laser at 785 nm. The bottom inset shows the trapping of peptides in the hotspot. (b) preprocessing techniques and preprocessed spectral maps (c) the sample spectra from each class and the highlights of the CNN architecture for peptide classifications, the detailed architecture of SAPNet Model provided in the supporting information.

**Methods**

**Materials**

Nonfunctionalized gold nanoparticles (AuNPs) with average particle size of 50 nm were from Sigma (753645-25ML, concentration of $3.5 \times 10^{10}$ particles/mL). The synthetic peptides were ordered from Biomatik. SYLGARD™ 184 Silicone Elastomer Kit was used for Polydimethylsiloxane (PDMS) microfluidic channel fabrication. Si wafers with 100nm $Si_3N_4$ membranes coated on the surface were purchased from MicroChemicals GmbH.

**Nanopore Fabrication**

The gold nanopores were fabricated on commercial $Si_3N_4$ membranes supported on silicon. The size and thickness of the $Si_3N_4$ window was 1 × 1mm and 100nm respectively. After sputtering a 2 nm titanium and 100 nm gold layer on the front side and 2nm gold layer at back side of the $Si_3N_4$ membrane, Focused Ion Beam (FIB) milling (FEI Helios DualBeam) from the back side of the membrane to create nanopores with 200nm diameter. Scanning electron microscope was used to characterize nanopore size and morphology. Then the nanopore samples were embedded in a microfluidic chamber made from PDMS.

**Attachment of amino acid and peptide on AuNPs**

All the amino acids and peptides used in the measurements were attached physically on gold nanoparticles. In the final solution for Raman measurements, the concentration of AuNPs and the salt concentration were $1.3 \times 10^{10}$ particles per mL and 5% of PBS buffer. Amino acid or peptide stock solutions were mixed with gold nanoparticles and PBS buffer. Before Raman measurement, the mixture were kept in a refrigerator at 4 °C for two days to allow the adsorption of analytes on AuNP. The concentrations of amino acids and peptides in the final solution were calculated according to previous literature to ensure the number of molecules adsorbed on each AuNP is far from forming a monolayer, therefore in the particle-in-pore system only one molecule occupies the hot spot and generates single molecule SERS signal. The details of the concentration calculation are in Supporting Information Table S1.

**Raman measurement of Single-molecule PTM**

We measured the SERS spectra of three amino acids and their PTMs (Proline, Hydroxyproline, Tyrosine, Phosphorylated Tyrosine, Serine, and Phosphorylated Serine) by using ThermoFisher DXR2xi Raman imaging microscope with a laser emitting at 785 nm, slip width of 50 μm, 15 mW laser power, 0.1 s exposure time. We also measured the SERS spectra of and two peptides and their PTMs (F-Ser, F-pSer, H-Pro and H-Hyp) whose sequences are shown in Table 1.

Table 1. The peptides and their sequences

| Peptides | Sequences | Origin of the peptides |
|---|---|---|

| F-Ser | Cys-Asp-Ser-Ser-Pro-Asp-Ser-Ala-Glu-Asp-Val-Arg-Lys | Alpha-2-HS-glycoprotein 2 (FETUA),[44] position: 132-144 |
|---|---|---|
| F-pSer | Cys-Asp-Ser-Ser-Pro-Asp-pSer-Ala-Glu-Asp-Val-Arg-Lys | |
| H-Pro | Leu-Glu-Met-Leu-Ala-Pro-Tyr-Ile-Pro | Hypoxia-inducible transcription factors (HIFs),[45] position 559-567 |
| H-Hyp | Leu-Glu-Met-Leu-Ala-Hyp-Tyr-Ile-Pro | |

As depicted in the sequence list in the table, F-Ser with F-pSer, and H-Pro with H-Hyp have quite similar chemical arrangment only differ by a modification in the single amino acid residue, with the modification occurring at one base in the peptides' PTM.

**SERS Signal Pre-processing**

Due to its suitability of detecting local patterns,[42,43] we chose a deep convolutional neural network (CNN). Before feeding the data into the CNN model, we conducted extensive SERS signal processing to ensure we have a consistent dataset (Figure 1b,c) We applied median filtering for cosmic ray removal (see supporting information Figure S1, S2). Subsequently, baseline correction is performed on the SERS spectra that are significantly affected by fluorescence-induced baseline during SERS measurement.[46,47] Because we measured the spectra in time series, during trapping events which generated strong SERS signal, the particle-in-pore system stayed in idling status and generate only noise basline, the most significant spectra were selected based on signal-to-noise ratio (SNR). A threshold of SNR = 2.5 is applied to select spectra correspoding to trapping event and preclude those belonging to system idling status. Figure 2(a) shows as an example of resulting data of hydroxyproline. Finally, the Savitzky-Golay [48] (S-G) smoothing algorithm and min-max normalization are applied to further remove noises from the selected spectra and normalize the SERS signals respectively. Figure 2b show typical single-molecule Raman spectra of the amino acids. Detailed implementation are available in the supporting information Figure S4, S5.

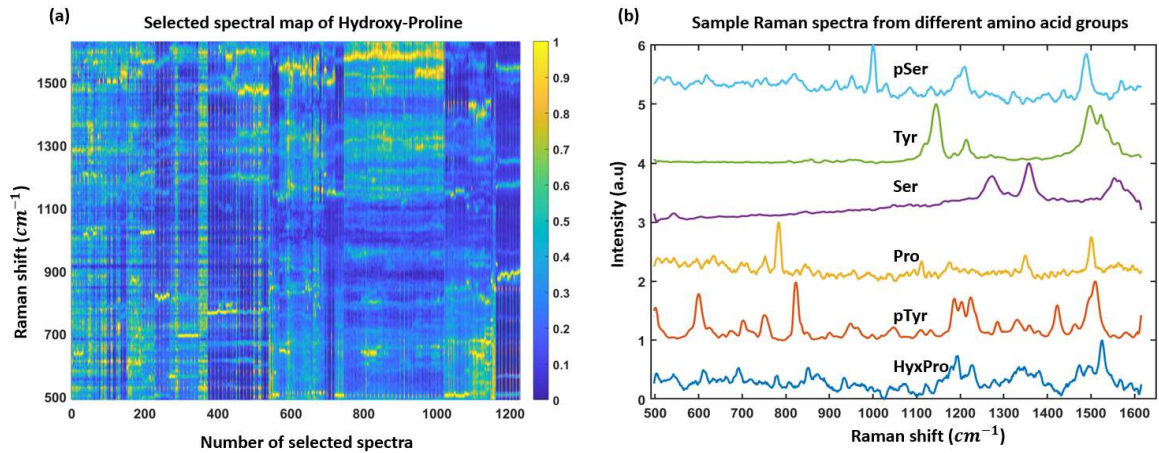

*Figure 2. (a) The spectral map of collected spectra of hydroxy proline; the spectra are selected from multiple measurements based on the SNR = 2.5 criterion and then combined together. (b) Sample spectra from different amino acids.*

**SAPNet model description**

After a series of SERS spectra preprocessing, we built the datasets having six different amino acids and four different peptides. Then we developed and validated a customized deep CNN to identify them. The architecture of SAPNet is illustrated in Figure 3, and the detailed implementation of SAPNet is also presented in the supplementary information Table S1. The network has an input, four convolutional blocks, a flattening layer, two fully connected blocks (dense blocks), and an output layer. Each convolutional block consists of two convolutional layers, two batch normalization (BN) layers, the maxpooling layer, and dropout layers. We started a convolutional layer with 64 filters and proceeded to a convolutional layer having 512 filters by doubling each time, which is the common practice in convolutional neural network.[49] The higher convolutional filters allow to extract abstract features from the input data. The importance of each max-pooling and dropout layer in each convolutional block is to reduce the spatial size and prevent overfitting, respectively.[50] The Batch normalization allows to fasten the training by introducing the higher learning rates and accelerating the convergence[51]. It also reduces the issues where the distribution of each layer's inputs during the training. The flattening layer[52] is used to transform multiple feature maps produced by the convolutional layers into 1D vector. The rectifier linear unit (ReLU) activation function introduces non-linearity to the model and mitigates the vanishing gradient.[53,54] The fully connected (Dense layer) basically do the final decision after the convolutional and pooling layers extract features, which should learn the global patterns in their input feature spaces to classify them. Finally, the softmax activation function in the output layer is used to convert raw score outputs into a probability distribution over multiple classes.

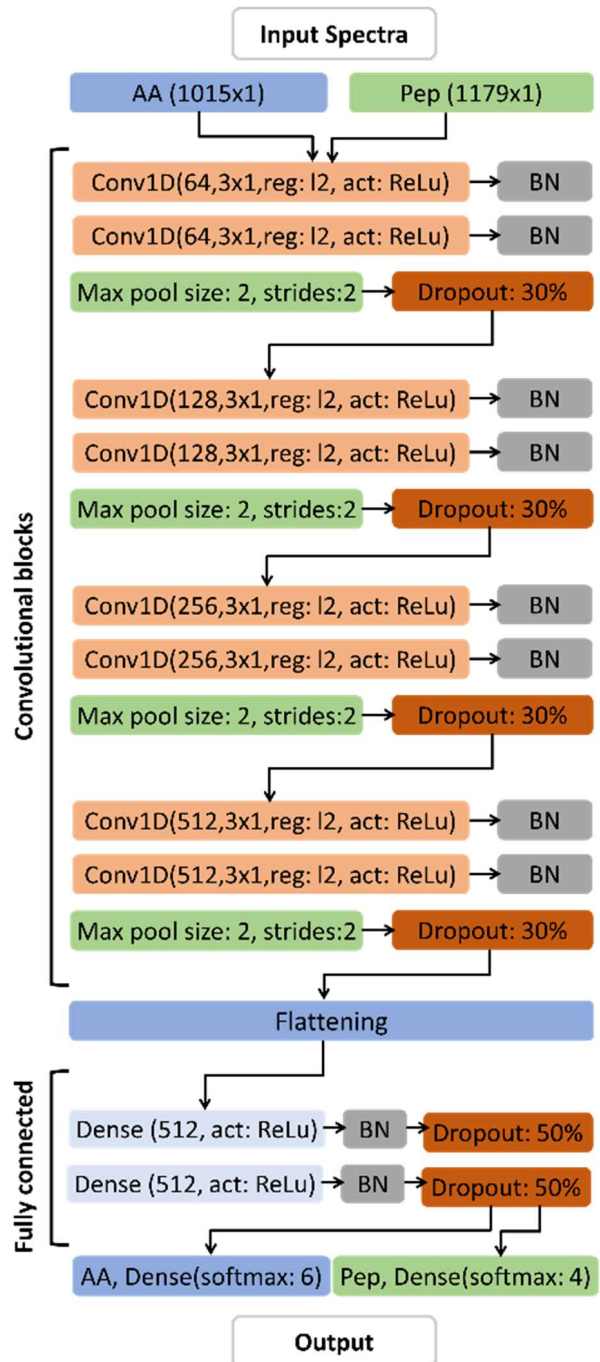

*Figure 3. The detailed convolutional neural network architecture of the SAPNet model. We used rectifier linear unit activation (ReLu) in each convolution block and SoftMax activation in the output layers. See the details in Table 1 in the supporting information.*

**Implementation Details and Model Evaluation**

The SAPNet model was adapted for the identification of amino acids and peptides with minor modifications. The amino acid (AA) dataset consists of 8,918 SERS spectra for training, validation and testing of the mode. Additionally, 469 AA spectra out of the training and testing datasets are used for the post-evaluation. Whereas the peptide dataset has 2 distinct peptides and their PTM with 3090 SERS spectra. Furthermore, we evaluated the model using 213 additional peptide SERS spectra. The dataset used to validate SAPNet is shown in Table 2. To develop and assess the model's performance, the original dataset was randomly divided into training and testing subsets in a 7: 3 proportion. Then the training dataset was further splitted into training and validation with a ratio of 4:1 respectively to evaluate the model during training.

Table 2. Amino acids and peptides dataset for validating the SAPNet model. The data does not include spectra used for out-of-sample-evaluation.

| **Amino Acids** | **Total** | **Training Set** | **Validation Set** | **Testing Set** |
|---|---|---|---|---|
| HyxPro | 1228 | 860 | 172 | 368 |
| PTyro | 2092 | 1464 | 293 | 628 |
| Pro | 1098 | 769 | 154 | 329 |
| Ser | 1480 | 1036 | 207 | 444 |
| Tyr | 1604 | 1123 | 225 | 481 |
| pSer | 1410 | 987 | 197 | 423 |
| **Peptides** | | | | |
| F-Ser | 571 | 400 | 80 | 171 |
| F-pSer | 1098 | 769 | 154 | 329 |
| H-Pro | 980 | 686 | 137 | 294 |
| H-Hyp | 441 | 309 | 62 | 132 |

We illustrated the performance of the model during training in Figure S6. (a and b) in the supporting information. The training and validation results align closely with each other. The SAPNet model was trained using an Adam optimizer, utilizing a kernel L2 regularization with a learning rate of 1e-4. Regularization helps in mitigating overfitting by penalizing overly complex models and encouraging simpler ones, thereby improving a deep learning model's ability to generalize to data points. [55,56] Adam optimizer is an adaptive gradient descent, which is well suited to solve complex optimization landscapes. Due to its computational efficiency and low memory requirements, using adaptive gradient algorithms will be beneficiary in analyzing Raman datasets. We utilized the standard Adam parameters of learning rate 0.0001, and cross-entropy loss to guide the model towards accurately classifying spectra. The performance metrics like loss, precision, AUC, and recall are calculated thoroughly to evaluate the performance of the model during the training phase. To prevent overfitting, dropout regularization of 30% and 50% were added after each convolutional block and dense layer respectively. Furthermore, early stopping was implemented by monitoring validation accuracy with patience of 30 epoches. This means that the model will terminate at its best performance if there is no improvement after 30 epochs. It also help mitigate overfitting. As shown in figure S8 in the supporting information, the loss is steadily decreasing during the training and the precision, area under curve and recall are increasing with epochs. A detailed description of the model architecture is presented to Table S2 in the supporting information.

**Results**

SAPNet model classified amino acids and peptides with a higher accuracy. The testing accuracy of the model is evaluated by the confusion matrix depicted in Figure 4(a) and (b). We achieved an overall accuracy of 99.66%, indicating the model successfully classifies amino acids from their modifications. To the right of the confusion matrix, we have calculated the true positive rate (TPR), and the false negative rate (FNR), which represent the correct or false classification of each class respectively. We got a high TPR and a very low FNR for each class, demonstrating the ability of the model to precisely recognize the target features from the Raman signature of amino acids. This quantitative evaluation confirms the model accurately identifies different classes of amino acids and peptides. In Figure 4(a) and 4(b), we presented the confusion matrices of SAPNet model for peptides, and amino acids classification, respectively. It provides a concise summary of the model's performance by depicting the distribution of predicted classes vs the actual classes.

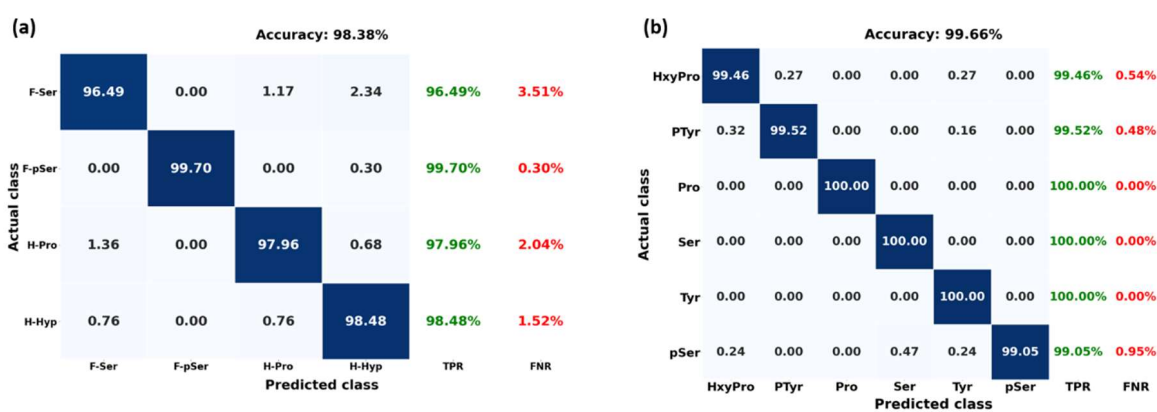

*Figure 4. The performance metrics of the SAPNet Model. The confusion matrice, True positive rate (TPR), and False positive rates of peptides and amino acids  4(a) and 4(b), respectively.*

To evaluate the overfitting, we demonstrated the prediction probability of unseen spectra based on the trained model in Figure 5. Evaluation of the model by unseen data during the training or testing demonstrates the robustness of the model. The model seems to be struggling more in identifying peptides from their PTM compared to recognizing amino acids from their modifications. One possible reason is that the residue with PTM may not in the hotspot at each trapping event that generate the SERS signal, considering the length of the peptides are longer than the hot spot size (covering 3 AA at most) [57,58] Another reason may be peptides share many amino acids in common.

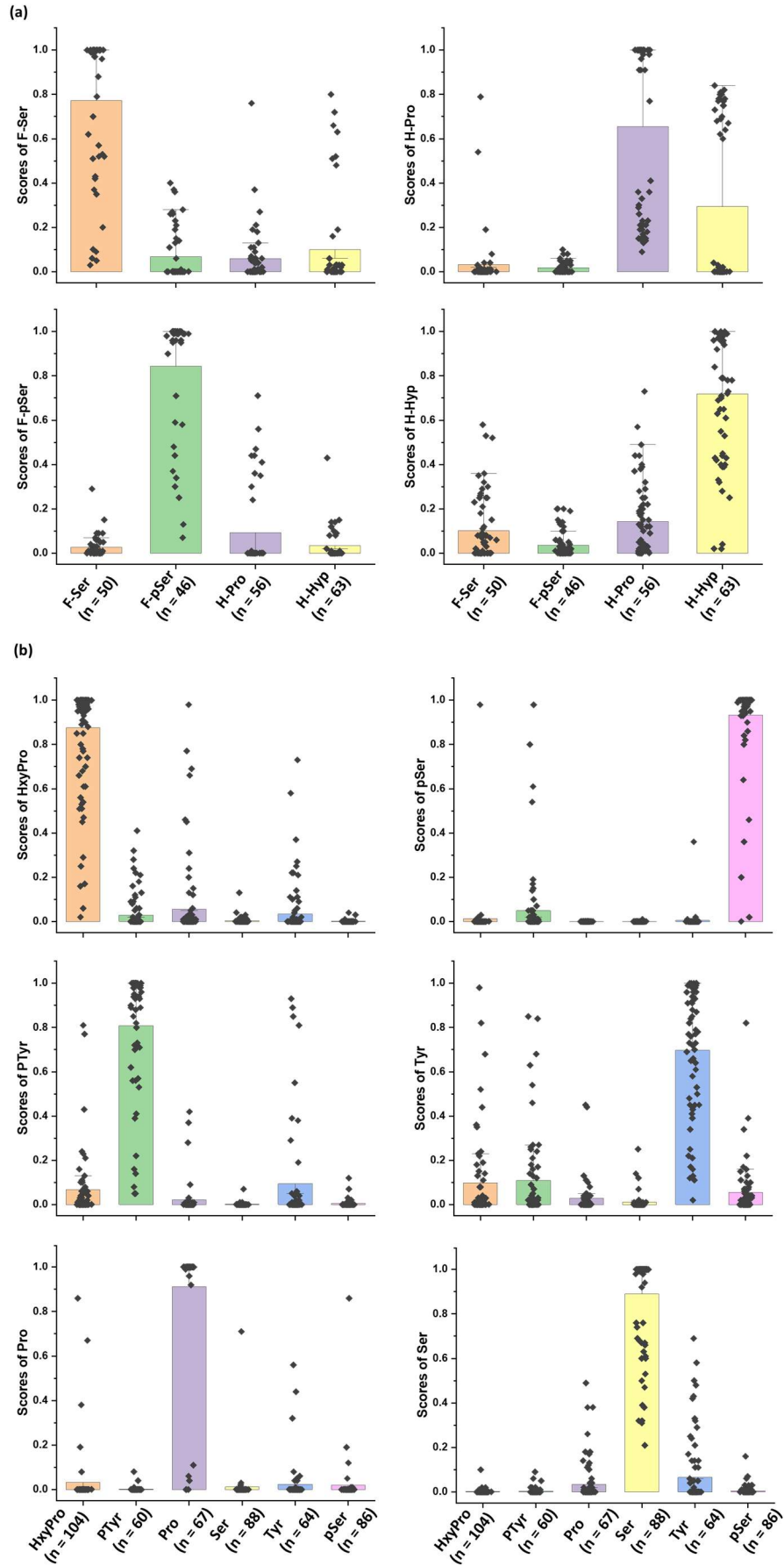

*Figure 5. The probability scores of unseen datasets on the SAPNet Model, including (a) peptide data, and (b) amino acid data, where n is the number of samples used to post-evaluate model. The square dots represent the probability values of each sample for different classes, with a higher prediction indicating the more likelihood of the sample belonging to the corresponding class.*

**Discussions**

We collected single molecule SERS signal of amino acid and peptide with phosphorylation and hydroxylation PTM. By using deep convolutional neural network SAPNet we able accurately identify both the amino acids and peptides with and without PTM. The model resulted in an excellent performance in classifying three amino acids and their counterparts with PTM. It also identifies the PTM in peptides H-Pro, H-Hyp, F-Ser and F-pSer derived from HIF-1α and EETUA. HIF-1α is known as a master regulator of oxygen homeostasis. These factors are particularly interesting in medical research due to their dual implications.[40]. The hydroxylation of proline residues in the degradation domains in the central region of HIF-1α will strongly change its affinity in the catalytic reaction.[59,60] Short peptide H-Pro had the same sequence of HIF-1α position 559-567, which contains one of the above-mentioned proline residues. H-Hyp had the hydroxylation modification of Pro at position 564. The second one is from the cistatin family, F-Ser FETUA, a multifunctional peptide responsible for several essential biological activities, but plays an important role in diabetes, kidney disease, and cancer. Due to its multiple and versatile presence in the human body, being able to observe and monitor this peptide is extremely beneficial in treatment and medical research related to tumor progression and sequelae of diabetes[41]. By measuring its PTM, disease-associated patterns could be extracted. For example, glycosylation and phosphorylation of the major hepatic plasma protein fetuin-A are associated with CNS inflammation in children.[61] Peptides F-Ser and F-pSer had the same sequence of FETUA position 132-144 and contain Ser and pSer respectively.[62] The successful discrimination of H-Pro, H-Hyp, F-Ser and F-pSer demonstrated the great potential of our technique in both mechanism studies as well as diagnosis.

In the post-evaluation, our models are evaluated by a new dataset distinct from the testing and training samples. This has two implications: firstly, it addresses the reproducibility of the model, and secondly, it shows the user interactivity of the models. The identification of amino acids and peptides at a single molecule level lays the foundation for further exploration into the integration of SERS with deep learning. Consequently, AI-powered spectroscopy will transform fields like biochemistry, molecular biology, proteomics, genomics, and medicine in the near future. Streamlining molecular characterization down to a single molecule level will have great promise for substantial progress in these scientific domains.

Our work suggests that deep learning is particularly well-suitable for single-molecule identifications. Analyzing thousands of SERS spectra by the human brain is time-consuming and prone to bias. Deep learning-based amino acid and peptide identification prove to be more efficient, saving time and avoiding human interventions. Our work achieved single molecule detection and identification from the mixture. The future work may address protein sequencing by combining SERS with hybrid deep neural network architectures.

**Author Information**

**Corresponding Authors:**

Jianan Huang, Email: jianan.huang@oulu.fi

Yingqi Zhao, Email: yingqi.zhao@oulu.fi

**Author contributions**

Mulusew Yaltaye determined the Raman spectra processing techniques, prepared the datasets, developed and validated the SAPNet model, and drafted the manuscripts. Yingqi Zhao designed the molecules, fabricated the Particle-in-Pore devices, determined the Raman measurement protocol, collected Raman spectra, and revised the manuscript. Eva Bozo contributed to the device fabrication, Raman spectra collection, and manuscript revision. Pei-Lin Xin helped in revising the manuscript. Francesco De Angelis contributed to the device fabrication and revised the manuscript. Vahid Farah revised the manuscript. Jianan Huang conceived the idea, supervised the work, and revised the manuscript.

**Acknolwdegement**

This research receives support from Academy Research Fellow project: TwoPoreProSeq (project number 347652), Biocenter Oulu emerging project (DigiRaman) and DigiHealth project (project number 326291), a strategic profiling project at the University of Oulu that is supported by the Academy of Finland and the University of Oulu.

**Additional information**

Supporting Information is available by request to the corresponding authors.

**References.**